\documentclass[12pt,preprint]{aastex}
\usepackage{mathrsfs}

\newcommand\beq{\begin{equation}}
\newcommand\eeq{\end{equation}}
\newcommand\bea{\begin{eqnarray}}
\newcommand\eea{\end{eqnarray}}
\newcommand\bal{\begin{align}}
\newcommand\eal{\end{align}}

\renewcommand\bal{\mbox{\boldmath$\alpha$}}

\begin{document}

\title{Hubble redshift and the Heisenberg frequency uncertainty: on
a coherence (or pulse) time signature in extragalactic light}

\author{Richard Lieu}
\affil{Department of Physics, University of Alabama, Huntsville, AL
35899.}

\begin{abstract}
In any Big Bang cosmology, the frequency $\omega$ of light detected
from a distant source is continuously and linearly changing (usually
redshifting) with elapsed observer's time $\delta t$, because of the
expanding Universe.  For small $\delta t$, however, the resulting
$\delta\omega$ shift lies beneath the Heisenberg frequency
uncertainty.  And since there {\it is} a way of telling whether such
short term shifts really exist, if the answer is affirmative we will
have a means of monitoring radiation to an accuracy level that
surpasses fundamental limitations.  More elaborately, had  $\omega$
been `frozen' for a minimum threshold interval before any redshift
could take place, i.e. the light propagated as a smooth but {\it
periodic} sequence of wave packets or pulses, and $\omega$ decreased
only from one pulse to the next, one would then be denied the above
forbiddingly precise information about frequency behavior. Yet
because this threshold period is {\it observable}, being e.g.
$\Delta t \sim$ 5 -- 15 minute for the cosmic microwave background
(CMB), we can indeed perform a check for consistency between the
Hubble Law and the Uncertainty Principle. If, as most would assume
to be the case, the former either takes effect without violating the
latter or not take effect at all, the presence of this
characteristic time signature (periodicity) $\Delta t$ would
represent direct verification of the redshift phenomenon.  The basic
formula for $\Delta t$ is $\Delta t \sim 1/\sqrt{|\alpha| \omega_0
H_0}$ where $H_0$ is the Hubble constant, $\omega_0$ is the mode
frequency at detection, and $\alpha=$ 1 for the cosmic microwave
background (CMB) and $\approx$ 0.1 for non-CMB extragalactic
sources. Thus, for the CMB one expects significant Fourier power,
that as given by the black body spectrum and no less, on the ten
minute timescale.  It is a clinching test.
\end{abstract}

\noindent

\noindent {\bf 1. Introduction}

The Hubble expansion, together with the CMB, remain to date the two
observational pillars of the $\Lambda$CDM standard Big Bang
cosmology.  With the advent of a remarkable modeling consistency of
both the CMB data and external correlations with other data (Spergel
et al 2007), the original elegant notion of baryonic matter and
radiation co-existing in a Universe governed by the theory of
General Relativity, which permits the presence of curvature in the
global geometry to bring about `finite but unbounded' space, was
dashed.  While striding success was also achieved in accounting for
structure formation from the acoustic oscillations of the CMB, the
ransom (see e.g. Brandenberger 2008, White 2007) is an avalanche of
extra assumptions: (a) dark matter, (b) dark energy, (c) inflation,
preceded by (d) quantum fluctuations in the matter density.  All
these, in addition to the earlier and historic conjectures of (e)
space expansion and (f) the Big Bang singularity, mean that
cosmologists no longer follow the long held astrophysics tradition
of using knowns to explain the unknown. Perhaps this is a healthy
transition, but even with all of (a) to (f) in full force, the
question of degeneracy comes next: the value of a key cosmological
parameter as inferred from the data depends on those of the other
parameters, yielding frequently to multiple best-fit solutions from
each cross-checking observation.

In this paper we suggest that it might be less presumptuous to take
matters one step at a time by querying whether, (a) to (f)
notwithstanding, just the Hubble expansion itself can be deduced
unambiguously from the properties of extragalactic radiation,
starting with the CMB. If space expands continuously, it should not
come as such a surprise to find evidence for this phenomenon in the
form of a unique time signature that radiation from remote sources
carries.  More precisely such radiation, unlike that emitted by
laboratory sources, has encoded in it the information of two vastly
different frequencies - the electromagnetic oscillation and the
Hubble constant.  The question not yet explored has to do with a
possible interplay between the two, that may lead to an intermediate
(or `beat') frequency more in tune with the tractable timescales of
our daily experience.

\vspace{2mm}

\noindent {\bf 2. CMB variability in an expanding Universe from
direct time domain analysis}

Let us first investigate the behavior of the most distant radiation,
the CMB, initially by considering one representative normal mode of
it, viz. an angular frequency of emission $\omega_e$ within the main
passband of the CMB black body spectrum.  As will be shown, the fact
that the full spectrum has a range of frequencies does not affect
the observational consequence to ensue from this treatment.

Unlike other cosmological sources, the CMB emission took place
throughout the entire Universe at a specific cosmological time
$\tau=\tau_e$, and the reason why we see a continuous signal around
the present epoch $\tau=\tau_0$ is because during earlier (later)
times $\tau < \tau_0$ ($\tau > \tau_0$) we detect this same CMB mode
as it was emitted at smaller (larger) comoving radii $r_e$ than the
radius of last scattering for $\tau=\tau_0$.

Thus, the complete wave phase of an evolving CMB mode may be written
as $\phi = \omega_e\eta - {\bf k_e} \cdot {\bf r} + \phi_e$, where
$\omega_e$ is the wave frequency at $\tau=\tau_e$ and $\eta$ is the
conformal time, defined as \beq \eta (\tau) = \int_{\tau_e}^\tau
\frac{d\tau'}{a(\tau')}, \eeq with $a(\tau)$ being the expansion
parameter at epoch $\tau$. For an observer ${\cal O}$ stationary (or
moving with velocity $v \ll c$) w.r.t. the cosmic substratum, who
performs measurements during some interval of local time $t$
centered at $\tau=\tau_0$, one can simply set ${\bf r} =0 $ in
$\phi$, by fixing the coordinate origin at the position of ${\cal
O}$.  Then the temporal part of the phase may be Taylor expanded
around $\tau_0$ to become \beq \phi(t) = \omega_0 t - \frac{1}{2}
\omega_0 H_0 t^2 + \phi_0, \eeq where $t=\tau - \tau_0$ and
$\omega_0 = \omega_e a_e/a_0 = \omega_e/(1+z)$.  Furthermore, the
CMB electric field amplitude scales with redshift as $(1+z)^2$, or
$(1-2H_0 t)$ for small $t$, because the CMB energy density $U \sim
|{\bf E}|^2$ has the redshift dependence of $U \sim (1+z)^4$. We may
now express the full expression of CMB electric field in the time
domain, as \beq {\bf E}(t)={\bf E_0} (1-2H_0 t) \exp \left[i\left(
\omega_0 t - \frac{1}{2} \omega_0 H_0 t^2 + \phi_0 \right)\right],
\eeq valid for times $H_0 |t| \ll$ 1.

Eqs. (2)  and (3) are interesting from one viewpoint: the angular
frequency of the mode, $\omega = \partial\phi/\partial t = \omega_0
(1-H_0 t)$, is monotonically decreasing with time.   Over an
interval $\delta t$ near the $t=0$ epoch, the frequency changes by
the amount $\delta\omega = -\omega_0 H_0 \delta t $, {\it however
small $\delta t$ and $\delta\omega$ may be}.  Is there a limit
beneath which such changes are merely theoretical?  By the
Uncertainty Principle the frequency parameter is for the same
$\delta t$ period a random variable (the wave is ambiguously defined
in finite time) with standard deviation $\sigma_\omega \gtrsim
1/\sigma_t \approx 1/\delta t$. In order to realize the redshift,
therefore, we must have $|\delta\omega| \gtrsim \sigma_\omega$. This
sets a threshold of $ |\delta t| \gtrsim \Delta t$ where \beq \Delta
t = \frac{1}{\sqrt{\omega_0 H_0}}, \eeq which is the {\it minimum
time} necessary for the expansion of the Universe to bring about a
physically consequential change in the CMB frequency.   In practice,
therefore, the CMB redshift is manifested as a sequence of `freeze
frames', each corresponding to a temporally coherent wave of finite
life, specifically a wave packet or {\it pulse} of constant $\omega$
(which is $\omega=\omega_0$ for the pulse centered at $t=0$) and
$\Delta t$ irrespective of whether the mathematical form of the\\
\newpage\noindent wave has on such scales an envelope or not\footnote{There
are two ways in which the CMB redshift can take place: (a)
continuously in extremely small steps, eq. (3), such that the wave
has no `envelope' on scales $> 2\pi/\omega_0$; (b) discretely in
steps of $\delta\omega \gtrsim \Delta\omega$ per $\delta t \gtrsim
\Delta t$ interval, in which case each transient frequency must be
`frozen into' a wave packet, or pulse, of size $\Delta t$, and the
separation between pulses is also $\Delta t$. Hence, if the CMB
shows no periodicity on the $\Delta t$ scale, then (unless we
abandon the expansion of space) scenario (a) will be the truth, and
we will have a means of {\it knowing for certain} that the CMB
frequency evolves systematically by amounts $|\delta\omega| \gg
\sigma_\omega$, which is just another way of saying that $\omega$
can be measured (or monitored) to an accuracy surpassing the
Uncertainty Principle.  In this way, (b) becomes the {\it only}
scenario that reconciles Hubble with Heisenberg.}.

From one CMB pulse to the next, not only is the coherence lost as
will be demonstrated below, but the light redshifts by the amount
\beq \Delta\omega = H_0 \omega_0 \Delta t \approx \sqrt{\omega_0
H_0}. \eeq These pulses define the {\it eigenmodes} that the
original black body mode of the CMB evolves into and out of, as the
Universe ages. If intermediate frequencies between the eigenvalues
are also used to describe the redshift, the effect will be
over-represented, essentially because the Uncertainty Principle
prevents the expansion from exerting a continuous influence on the
CMB (or any propagating radiation). Owing to the discreteness (or
quantization) of the redshift process, the CMB signal for this mode
should then exhibit a periodic time signature, with successive
pulses marking the times of maximum photon arrival rate separated by
one $\Delta t$ interval.

To examine more closely the consequences of eqs. (3) and (4), let us
return to the electric field ${\bf E}(t)$ of the CMB mode, eq. (2).
Now the phase of ${\bf E}(t)$ is coherent only when $|t| \lesssim
\Delta t$, in the sense that when $|t| \gtrsim \Delta t$ the phase
of the wave is {\it not} at the value as given by the constant
frequency expectation of $\omega_0 t + \phi_0$, but differs from it
by an amount $\delta\phi \approx 2\pi$, i.e. by the time $t$ is so
far ahead (or behind) our time origin of $t=0$, the wave phase has
evolved `out of step' by one full wave cycle, or a substantial
portion thereof, and it is no longer possible to treat the frequency
as $\omega=\omega_0$ for these times.  Hence, as will also be
demonstrated in detail by the technique of Fourier transform, the
segment of coherent wave is centered at $t=0$, of lifetime $\Delta
t$ and having a wave frequency $\omega_0$.   At times $t > \Delta t$
the mode is `phase locked' to another wave, of frequency $\omega =
\omega'_0 \approx \omega_0 - \Delta\omega$, for a further interval
of time $\Delta t' = 2\pi/\sqrt{\omega'_0 H'_0} \approx \Delta t$,
which defines the next segment.  This is compelling evidence for the
Uncertainty Principle: within a segment of coherence the frequency
must be regarded as frozen, as there is no physical effect of any
kind to suggest otherwise.  It is therefore meaningless to
contemplate frequency changes ensuing from eq. (3) over a shorter
elapsed time than $\Delta t$.

\vspace{2mm}

\noindent {\bf 3.  CMB time signature by Fourier transform}

In order to confirm the heuristic analysis of the previous section,
and to see how a CMB black body mode evolves into modes of lowering
frequencies as time progresses, it is necessary to enter Fourier
space.  By eq. (3) the Fourier amplitude is \beq {\bf E} (\omega) =
\int_{-T/2}^{T/2} {\bf E_0} (1-2H_0 t) \exp -i\left[(\omega -
\omega_0) t + \frac{1}{2} \omega_0 H_0 t^2 -\phi_0 \right] dt. \eeq

It is possible to find the primitive function of the indefinite
integral, viz.

\begin{eqnarray}
{\bf E}(\omega,t) = & {\bf E_0} &(1-i) \sqrt{\pi}
\omega_0^{-\frac{3}{2}} H_0^{-\frac{1}{2}} \left(\omega -
\frac{\omega_0}{2}\right) H_0 e^{-\frac{i(\omega -
\omega_0)^2}{2\omega_0 H_0}} {\rm erfi} \left[\frac{(1+i)(\omega -
\omega_0 + \omega_0 H_0 t)}{2\sqrt{\omega_0 H_0}}\right] \nonumber \\
&+&  \frac{2i}{\omega_0} \exp\left[i(\omega - \omega_0) t +
\frac{i\omega_0 H_0 t^2}{2}\right],
\end{eqnarray}
where erfi$(z)$ is the imaginary error function, defined as \beq
{\rm erfi} (z) = \frac{2}{\sqrt{\pi}} \int_0^z e^{\xi^2} d\xi, \eeq
where we ignored the epoch phase $\phi_0$ in eq. (6).  The last term
on the right side of eq. (7) can be neglected, because with
$\omega_0 \gg H_0$ it's amplitude is much less than that of the
preceding term.

We can now look at the time intervals within which CMB radiation at
various frequencies $\omega$ arrive, starting with
$\omega=\omega_0$.  By means of the approximation ${\rm erfi}(z)
\approx 2z/\sqrt{\pi}$ for $|z| \ll$ 1, where $z=
(1+i)\sqrt{\omega_0 H_0}T/4$, we see that $|{\bf E}(\omega_0)|$
rises linearly with $T$ when $T \ll \Delta t$, with $\Delta t$ as
given by eq. (4). This is no different from the behavior of the
spectral amplitude at $\omega=\omega_0$ when an infinite
monochromatic wave with frequency $\omega_0$ is sampled for a time
$T$.  As $T$ continues to increase, however, ${\rm erfi}(z)$ will
turn over.  At large $|T|$ the erfi function assumes the asymptotic
form \beq {\rm erfi} (z) \approx \frac{e^{z^2}}{\sqrt{\pi} z} +
\frac{z}{\sqrt{-z^2}}, \eeq enabling us to write \beq {\rm erfi}
\left[\frac{(1+i)\sqrt{\omega_0 H_0}t}{2}\right] \approx
\frac{2}{(1+i)\sqrt{\pi\omega_0 H_0} t} e^{\frac{i\omega_0 H_0
t^2}{2}} \pm i~~{\rm at}~t=\pm \frac{T}{2} \eeq where the $\pm$ sign
refers to $t > 0$ and $t<0$ respectively.

Thus, when $T \rightarrow \Delta t$, ${\bf E}(\omega)$ {\it
saturates} to the value of \beq {\bf E}(\omega) = {\bf E_0}
2\sqrt{\pi}(1+i) \omega_0^{-\frac{3}{2}}
H_0^{-\frac{1}{2}}\left(\omega - \frac{\omega_0}{2}\right)
e^{-\frac{i(\omega - \omega_0)^2}{2\omega_0 H_0}}. \eeq  An
immediate test of the validity of our calculation is afforded by
letting $H_0 \rightarrow 0$, in which case $\Delta t \rightarrow
\infty$ (hence $T$ also) and one must recover the expected Fourier
transform of an infinite plane wave. Indeed as $H_0 \rightarrow 0$
we have \beq {\bf E}(\omega) = {\bf E_0} (1+i)
\sqrt{\frac{\pi}{\omega_0 H_0}} e^{-\frac{i(\omega -
\omega_0)^2}{2\omega_0 H_0}} \rightarrow 2\pi \delta (\omega -
\omega_0), \eeq where in the last step use was made of the relation
$$ \lim_{s\rightarrow 0^+} (1-i)\sqrt{\frac{\pi}{s}}
\exp\left(-\frac{ix^2}{2s}\right)=2\pi\delta (x), $$ which is a
standard limiting form of the Dirac delta function\footnote{See e.g.
\texttt{http://functions.wolfram.com/GeneralizedFunctions/DiracDelta/09/0005/.}}.

Eq. (11) is clear indication that the CMB mode does not switch
frequency from one instance to another. Rather, there exists a
finite interval $T \sim \Delta t$, the coherence length of section
2, during which the mode evolves into and out of the frequency
$\omega=\omega_0$. Moreover, as a self-consistency check we find
that for other `core' frequencies $\omega \neq \omega_0$ but \beq
|\omega - \omega_0| \lesssim \Delta\omega, \eeq with $\Delta\omega$
as defined in eq. (5), the linear increase of ${\bf E}(\omega)$
towards its maximum of eq. (11) still holds, i.e. the persistence of
the $\omega=\omega_0$ wave for a finite time $\Delta t$ leads to a
spreading of the spectral line by $\Delta\omega$ in accordance with
the Fourier bandwidth theorem. As a result, the frequency components
within this line are coherent. It is therefore the correct procedure
to model the evolution of a CMB mode only in terms of a set of
uncorrelated eigenfrequencies separated by $\Delta\omega$, so that
the lines do not overlap; nor do the corresponding pulses in the
time domain, which are resolved by one spacing of $\Delta t$.

To appreciate further the importance of coherence, let us now turn
to the `wing' frequencies, viz. those with $|\omega-\omega_0|
\gtrsim \Delta\omega $.  Here, the asymptotic formula for ${\rm
erfi} (z)$ becomes \beq {\rm erfi} \left[\frac{(1+i)
(\omega-\omega_0 + \omega_0 H_0 t)}{2\sqrt{\omega_0 H_0}}\right]
\approx \frac{2\sqrt{\omega_0 H_0}}{(1+i)\sqrt{\pi} (\omega -
\omega_0)} e^{\frac{i(\omega - \omega_0)^2}{2\omega_0 H_0}}
e^{i(\omega-\omega_0)t}. \eeq  After inserting the integration
limits $t=\pm T/2$, one  sees that at this `wing' region ${\bf
E}(\omega)$ defaults to the familiar `sinc function', viz. \beq {\bf
E}(\omega) = -{\bf E_0}
\frac{i\sin\left[\frac{1}{2}\left(\omega-\omega_0\right)T\right]}{\omega-\omega_0}.
\eeq  So the overall conclusion is that, so long as $T \lesssim
\Delta t$, both at the `core' and `wing' the spectrum resembles
closely that of a $\omega=\omega_0$ wave when the wave is observed
for the duration $T$.  Once $T > \Delta t$, the central amplitude
saturates at the value of eq. (11) and no longer grows with $T$ like
a `sinc' function at $\omega \approx \omega_0$ does.  Instead, the
width of the `core' continues to enlarge at constant height, to
exceed $\Delta\omega$ (i.e. more and more of the `wing' amplitudes
are lifted from their values in eq. (14) to those in eq. (12)), due
to the linear superposition of line profiles from other
eigenfrequencies that now begin to assume importance as the
$\omega=\omega_0$ wave loses coherence. The entire situation is
depicted in Figure 1.

Turning to the broad band nature of the CMB black body spectrum,
which consists of many normal modes with random relative phases at
emission, during the observation each redshifting black body mode
will `stampede' across the passband of the observer's telescope
filter, as it triggers repetitive pulses of microwave energy, all at
approximately the period $\Delta t$.   Hence the measured CMB
intensity profile is the superposition of many pulses with different
phases, and periods $\Delta t$ ranging\footnote{At the peak of the
black body spectrum $\omega = \omega_0^{{\rm max}} \approx $
10$^{12}$ Hz ($\nu_0^{{\rm max}} \approx$ 160.4 GHz) we have from
eq. (4) $\Delta t \approx$ 675 s; then the scaling of $\Delta t \sim
1/\sqrt{\omega_0}$ and the frequency spread among modes of $\delta
\omega \approx \omega_0^{{\rm max}}$  can be used to estimate the
pulse width dispersion.} from $\sim$ 540 s to $\sim$ 945 s.

Thus the output telescope signal will exhibit a long term behavior
that is aperiodic, yet containing a {\it significant excess} of
Fourier power on the hourly scales, even though the domination of
noise over these scales may present formidable challenges to any
data processing and interpretation effort.  Specifically the power
for each timescale is given by the black body function $B(\omega)$
at the corresponding mode.  In this way a single model involving
$B(\omega)$, the telescope response function, and $H_0$ as the only
free parameter, can be employed to fit the FFT data stream of an
appropriate CMB observation.

\vspace{2mm}
%\newpage

\noindent {\bf 4.  Time signature of non-CMB extragalactic sources}

For a non CMB extragalactic source that does not have a large
peculiar velocity, let one radiation mode of definite frequency
$\omega_e$ be emitted at epoch $\tau=\tau_e$, to arrive at $r=0$ for
reception at epoch $\tau=\tau_0$.  If more photons are detected at
later epochs $\tau>\tau_0$, they would have been emitted at $\tau >
\tau_e$ from the {\it same} comoving distance $r$.  This property
distinguishes all other sources from the CMB: it is the spread of
emission time rather than source distance that leads to a continuous
light signal.  Returning to the arguments of section 2, the
constancy of $r$ sets the constraint \beq \int_{\tau_e +
\delta\tau_e}^{\tau_0 + \delta\tau_0} \frac{d\tau}{a(\tau)} =
\int_{\tau_e}^{\tau_0} \frac{d\tau}{a(\tau)}, \eeq which simplifies
to $\delta\tau_e/a_e = \delta\tau_0/a_0$, or $\delta t_e = \delta
t_0/(1+z)$ where once again we expressed time changes in terms of
the local clock $t$.   Thus, if the frequency detected at $t=0$
($\tau=\tau_0$) is $\omega = \omega_0$, its value at $t=\delta t_0$
will be \beq \omega'_0 = \omega_e \frac{a_e + \dot a_e \delta
t_e}{a_0 + \dot a_0 \delta t_0} = \omega_0 \frac{1+H_e\delta
t_e}{1+H_0 \delta t_0}, \eeq where $H_e=\dot a_e/a_e$ is the Hubble
constant at $\tau=\tau_e$, which is related to $H_0$ by the standard
formula \beq H_e = H_0 E(z) = H_0 [\Omega_m (1+z)^3 +
\Omega_{\Lambda}]^{\frac{1}{2}}, \eeq where $\Omega_m \approx$ 0.3
and $\Omega_{\Lambda} \approx$ 0.7 for $\Lambda$CDM (Bennett et al
2003, Spergel et al 2007).

Proceeding from eq.  (17) a little further, one readily obtains the
phase of the arriving wave as $\phi= \omega_0 t(1-H_0\alpha t/2)
+\phi_0$ where $\alpha = \alpha (z) = 1 - E(z)/(1+z)$.  Hence the
`beat' period for light from an extragalactic source at redshift $z$
is $\Delta t \approx 1/\sqrt{|\alpha|\omega_0 H_0}$.  By means of
eq. (18), one finds a typical value for $|\alpha|$ among a wide
redshift range between $z=$ 0.2 and $z=$ 4.0 of $|\alpha| \approx$
0.1.   Hence a representative $\Delta t$ for optical and UV sources
with $\omega_0 \approx$ 10$^{15}$ Hz will be $\Delta t \approx$ 67.5
s, or one minute.

\vspace{2mm}

\noindent {\bf 5. Conclusion}

The continuous redshifting of light from extragalactic sources sets
a temporal coherence time, or pulse time which is a scale of immense
importance in physical optics, to the arriving signals.  One should
therefore expect the effect to be observable, particularly in the
form of quasi-periodic peaks separated by $\sim$ ten minute for the
CMB, and $\sim$ one minute for optical sources.  This would be a
direct verification of the Hubble expansion phenomenon.

One possible way of carrying out such a measurement is to look at a
fixed area of the last scattering surface for a period of $\sim$
hours to two days and with a resolution $\sim$ a minute, and analyze
the data by e.g. FFT. It is not the purpose of this work, however,
to find out whether past and present missions (like COBE/FIRAS,
Mather et al 1994; and WMAP, Bennett et al 2003, Spergel et al 2007)
have already gathered the necessary data to deliver a verdict, nor
what future planned missions (most notably Planck, Tauber et al
2003) can accomplish within the scope of their existing
observational timeline.

%The author thanks Prof. James Miller of Huntsville, and Lord James
%McKenzie of the Hebrides and Outer Isles for helpful discussions.

%\vspace{3mm}
\newpage

\noindent {\bf 6. References}

\noindent Bennett, C.L. et al 2003, ApJS, 148, 1.

%\noindent Bielby, R., Shanks, T., 2007, MNRAS, 382, 1196.

\noindent Brandenberger, R. 2008, Physics Today, 61, 44.

%\noindent Durret, F., Kaastra, J.S., Nevalainen, J., Ohashi, T.,\\
%\indent Werner, N., 2008, Sp. Sci. Rev. in press (arXiv:0801.0977).

%\noindent Goulielmakis et al 2004, Science, 305, 1267.

%\noindent Gunn, J.E., and Ostriker, J.P., 1971, ApJ, 165, 523.

%\noindent Kaastra, J.S. et al 2008, Sp. Sci. Rev. in press
%(arXiv:0801.0964).

%\noindent Lieu, R., Mittaz, J.P.D., Zhang, S.N., 2006, ApJ, 648,
%176.

%\noindent Lighthill, M.J., 1978, Waves in Fluids, Cambridge
%University Press.

\noindent Mather, J.C. et al 1994, ApJ, 420, 439.

\noindent Spergel, D.N. et al 2007, ApJS, 170, 377.

\noindent Tauber, J.A. et al 2003, Advances in Space Research, 34,
491.

\noindent White, S.D.M., 2007, Rep. Prog. Phys., 70, 883.

\begin{figure}
\vspace{-10cm}
\begin{center}
\hspace{-2cm}
\includegraphics[angle=0,width=7in]{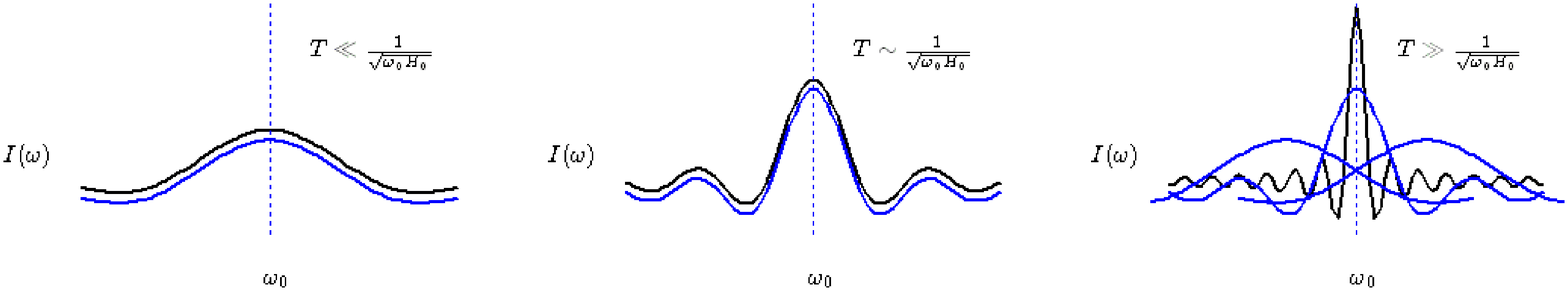}
\vspace{-1cm}
\end{center}
\caption{Spectral dependence of one CMB black body mode on sampling
interval $T$ (blue line),  with $\omega_0$ as the epoch frequency,
as compared with that of an $\omega=\omega_0$  infinite plane wave
(black line, the `sinc' function).  To comply with the Uncertainty
Principle, the two lines follow each other closely until the the CMB
mode loses coherence with the plane wave. Beyond that (rightmost
graph) the two spectra evolve separately: the central blue curve
saturates in height as the Hubble expansion brings other eigenmodes
to existence, in the form of further pulses which are part of a
periodic sequence.  At this point the total CMB spectrum is the sum
of all the blue curves.}
\end{figure}

\end{document}